%% file: MgH2.tex
\documentclass[10pt]{iopart}
\newcommand{\IMSS}{Muon Science Laboratory and Condensed Matter Research Center, Institute of Materials Structure Science, High Energy Accelerator Research Organization (KEK-IMSS), Tsukuba, Ibaraki 305-0801, Japan}
\newcommand{\Sokendai}{Department of Materials Structure Science, The Graduate University for Advanced Studies (Sokendai), Tsukuba, Ibaraki 305-0801, Japan}
\newcommand{\Ibadai}{Institute of Quantum Beam Science, Graduate School of Science and Engineering, Ibaraki University, Mito, Ibaraki 310-8512, Japan}
\newcommand{\IMR}{Institute for Materials Research, Tohoku University, Sendai 980-8577, Japan}
\newcommand{\JAEA}
{Advanced Science Research Center, Japan Atomic Energy Agency, Tokai, Ibaraki 319-1195, Japan}

\usepackage{iopams}
\usepackage[dvipdfmx]{graphicx}
\usepackage{dcolumn}
\usepackage{multirow}
\usepackage{bm}
\usepackage[hypertex,colorlinks=true,linkcolor=black,citecolor=blue,filecolor=blue,urlcolor=blue,setpagesize=false,nesting=true]{hyperref}

\usepackage[utf8]{inputenc}
\usepackage[T1]{fontenc}
\usepackage{mathptmx}

\newcommand{\msr}{$\mu$SR}%

\begin{document}
\title{Local electronic structure of interstitial hydrogen in MgH$_2$ inferred from muon study}

\author{Ryosuke Kadono$^1$, Masatoshi Hiraishi$^2$, Hirotaka Okabe$^3$,\\ Akihiro Koda$^{1,4}$, and Takashi U Ito$^5$}

\address{$^1$ \IMSS}
\address{$^2$ \Ibadai}
\address{$^3$ \IMR}
\address{$^4$ \Sokendai}
\address{$^5$ \JAEA}
\ead{ryosuke.kadono@kek.jp}

\begin{abstract}
Magnesium hydride has great potential as a solid hydrogen (H) storage material because of its high H storage capacity of 7.6 wt\%. However, its slow hydrogenation and dehydrogenation kinetics and the high temperature of 300 $^\circ$C required for decomposition are major obstacles to small-scale applications such as automobiles. The local electronic structure of interstitial H in MgH$_2$ is an important fundamental knowledge in solving this problem, which has been studied mainly based on density functional theory (DFT). However, few experimental studies have been performed to assess the results of DFT calculations. We have therefore introduced muon (Mu) as pseudo-H into MgH$_2$ and investigated the corresponding interstitial H states by analyzing their electronic and dynamical properties in detail.  As a result, we observed multiple Mu states similar to those observed in wide-gap oxides, and found that their electronic states can be attributed to relaxed-excited states associated with donor/acceptor levels predicted by the recently proposed ``ambipolarity model''.  This provides an indirect support for the DFT calculations on which the model is based via the donor/acceptor levels.  An important implication of the muon results for improved hydrogen kinetics is that dehydrogenation, serving as a {\sl reduction} for hydrides, stabilises the interstitial H$^-$ state.
\end{abstract}
\noindent{\it Keywords:} hydrogen strage, interstitial hydrogen, electronic structure, muon spin rotation
\maketitle
\ioptwocol
\section{Introduction}
Hydrogen (H) storage materials are attracting attention for their importance as a core energy carrier in hydrogen energy production and utilization systems toward a decarbonized society \cite{Dresselhaus:01,Schlapbach:01}. Therefore, there is a need for materials that can safely and efficiently store H and easily control the absorption and release of H. In particular, solid materials such as metal hydrides are characterized by high safety and high volume and weight densities, and there are great expectations for their practical use \cite{Sakintuna:07}.

Among metal hydrides, magnesium hydrides have always attracted considerable interest. Magnesium reacts reversibly with H to form hydride (MgH$_2$), which has great potential as a solid H storage material because of its high H storage capacity of 7.6 wt\%. However, its slow hydrogenation and dehydrogenation rates and the high temperature of 300 $^\circ$C required for decomposition are major obstacles to small-scale applications such as automobiles \cite{Sakintuna:07,Felderhoff:09}. Several experimental and theoretical studies have already been published on H kinetics in MgH$_2$ to solve the problem. However, most of them have concentrated only on macroscopic analysis by thermodynamic measurements of the effects of ball milling, heavy ion irradiation, or the addition of trace metals or their oxides on the reaction kinetics
\cite{Huot:99,Oelerich:01,Fernandez:02,Liang:04,Conradi:07,Du:08,Evard:10}.

Understanding the atomic-level mechanisms of hydrogenation and dehydrogenation processes is key to engineering solutions for accelerating these processes and lowering the decomposition temperatures. In MgH$_2$, the local electronic state of isolated H at the interstitial site (H$_{\rm i}$) is also one of the most fundamental pieces of information in elucidating the rate-limiting processes in H kinetics. Meanwhile, experimental means to study the microscopic states of isolated H in trace amounts in solids are limited, and so far the {\it ab initio} calculations based on density functional theory (DFT) have been the primary tool for this purpose \cite{Du:08,Park:09,Roy:13}. 
However, there have been few experimental studies to assess the theoretical predictions to date. 

To overcome this situation, we have introduced positive muon ($\mu^+$) into MgH$_2$, and investigated their electronic states and dynamical properties in detail using muon spin rotation ($\mu$SR).  Because the mass of $\mu^+$ is two orders of magnitude (about 206 times) greater than the electron mass, the adiabatic approximation is sufficient for understanding $\mu^+$-electron interaction. In fact, the difference in the Bohr radius between a $\mu^+$ binding a single electron, known as muonium, and the corresponding H$^0$ atom in vacuum is only 0.43\%, implying that they have nearly the same electronic structure in the matter.  Hereafter, the elemental symbol Mu will be introduced to denote $\mu^+$ as pseudo-H, and the valence electronic states of Mu will be denoted as Mu$^+$, Mu$^0$, and Mu$^-$. For deliberately avoiding distinction between Mu$^+$ and Mu$^-$, both are referred to together as ``diamagnetic Mu'' and Mu$^0$ is called ``paramagnetic Mu''.

In comparing experimental and theoretical results, there is an important issue to be pointed out at this stage. In conventional $\mu$SR experiments, $\mu^+$s are implanted as a relatively high-energy ion beam (typical kinetic energy of $\sim$4 MeV), which generates free carriers and excitons in insulator crystals due to the associated electronic excitation towards the end of the radiation track \cite{Thompson:74,Alig:75,Itoh:97}. These often propagate rapidly in the crystal, and there is experimental evidence that Mu acts as a capture center to form relaxed-excited states upon interaction with free carriers and/or excitons \cite{Hiraishi:22}.  Accordingly, these Mu states do not necessarily correspond to the electronic state predicted from the thermodynamic charge conversion level ($E^{+/-}$) obtained by DFT calculations for thermal equilibrium conditions. Moreover, Mu is often observed to be accompanied by fast spin relaxation, which is due to spin/charge exchange interactions with excited carriers and/or excitons. While these phenomena are not observed for H in thermal equilibrium without such carriers, they can be regarded to simulate H in bulk excited states induced by light or electric fields, and Mu is useful as a microscopic probe for such states.

Another well-known evidence that Mu exhibits relaxed-excited states is that two or more different Mu electronic states are often observed simultaneously in the same material.  In particular, two paramagnetic Mu states, the tetrahedral center Mu$^0$ and the bond center Mu$^0$, have long been known in elemental semiconductors and III-V compound semiconductors \cite{Patterson:88}, and these have been attributed to the states associated with the temporal acceptor and donor levels ($E^{0/-}$ and  $E^{+/0}$) rather than $E^{+/-}$ \cite{Lichti:08}. With this in mind, we have recently focused on the diamagnetic Mu states, which are  observed in pairs with Mu$^0$ in many wide-gap oxides but whose origin has been unexplored, and conducted an extensive survey to compare the previous experimental results and DFT calculations in the literature. As a result, assuming that the diamagnetic Mu corresponds to the donor-like state which is ionised because the $E^{+/0}$ level is located within the conduction band, we found that the valence of the paired Mu states can be predicted from the positions of $E^{0/-}$ and $E^{+/0}$ in the band structure \cite{Hiraishi:22}.   This allows us to describe the previous experimental results in a unified manner, and we call it the ``ambipolarity model'' because it is ascribed to the fact that the ambipolarity, which is the true nature of H, is expressed through these relaxed-excited states.

MgH$_2$ crystalises in a stable rutile structure with few H vacancies at ambient condition (the decomposition reaction is endothermic \cite{Felderhoff:09}). It is a typical ionic insulator with a band gap above 5 eV, and the ambipolarity model combined with the results of DFT calculations for H predicts that Mu simultaneously takes on both acceptor and donor-like electronic states. The present experimental results are consistent with this prediction, suggesting that the ambipolarity model holds semi-quantitatively in MgH$_2$. In the following, we will therefore organise the experimental results in terms of this model and attempt to interpret them based on their consistency with the predictions from the corresponding DFT calculations.

\section{$\mu$SR Experiment and DFT Calculations}
The sample used for $\mu$SR measurements was obtained from a commercial vendor (Fujifilm Wako Chmicals Co.) as a powdered reagent sealed with inert gas.  Immediately after opening the container, approximately 2 g of the powder was pressurised and shaped into a disk covered with aluminum foil, mounted on a temperature-controlled cryostat, and then degassed into a vacuum. The $\mu$SR experiments were performed using the ARTEMIS spectrometer installed in the S1 area at the Materials and Life Science Experimental Facility in J-PARC, where a nearly 100\% spin-polarised pulsed muon beam (25 Hz, with the full width at half-maximum of 80 ns and a momentum of 27 MeV/c) was transported to the sample.  $\mu$SR spectra [time-dependent positron asymmetry, $A(t)$] were measured from room temperature to 5 K in zero magnetic field (ZF)  and under longitudinal magnetic fields (LF, 0.01--0.4 T).

The donor/acceptor levels for H assessed in this work are those based on Ref.~\cite{Roy:13}. They performed DFT calculations with screened hybrid functions of Heyd, Scuseria, and Ernzerhof (HSE) as implemented in the Vienna {\it Ab initio} Simulation Package (VASP) code.  The amount of exact exchange mixed into the HSE generalized gradient functional is taken to 33\%, which reproduces the experimentally observed lower limit of the band gap (5.2 eV). Defect calculations were performed for a supercell containing 72 atoms with $2\times2\times3$ iterations of the primitive cell. A mesh of $2\times2\times2$ special ${\bm k}$ points and an energy cutoff value of 270 eV for the plane-wave basis set were used for the defective supercells. For charged defects, the number of electrons in the supercell was varied according to the format described in Ref.~\cite{Walle:04} and the electrons were localized on the impurity.

\begin{figure}[t]
\begin{center}
\includegraphics[width=0.7\linewidth]{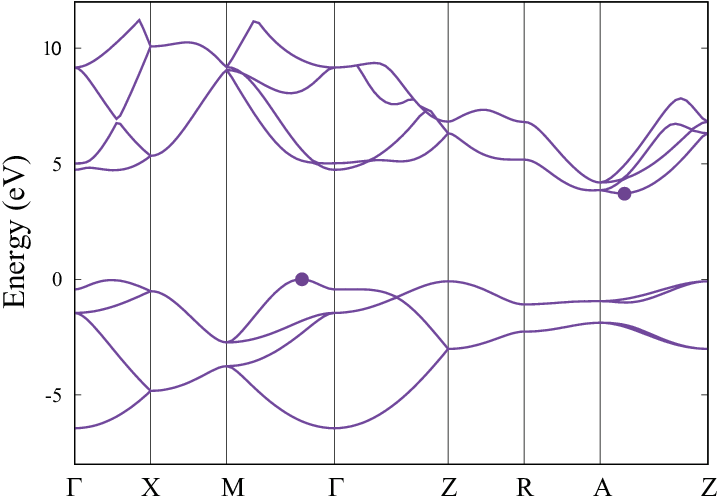}
\caption{Band structure of MgH$_2$ calculated by QUANTUM ESPRESSO code (see text). Filled circles indicate the conduction band minimum  and valence band maximum.}
\label{band}
\end{center}
\end{figure}

We have also performed DFT calculations to investigate in detail the local structure of H-related defects and to evaluate hyperfine parameters of Mu$^0$ and $\Delta$ of the diamagnetic Mu; the comparison of these with experimental values allows us to identify the corresponding donor/acceptor states involving Mu. The DFT calculations were within the generalized gradient approximation (GGA) and the projector augmented wave method as implemented in the QUANTUM ESPRESSO (QE) code \cite{Giannozzi:09}. The defects were simulated using 48-atom host supercell with 2$\times$2$\times$2 iterations of the primitive cell, and the Brillouin-zone integrations implemented by summations over a Monkhorst-Pack $k$-point mesh of 3$\times$3$\times$5, with an energy cutoff of 100 Ry for the plane-wave basis set.
The calculated lattice parameters $c=0.301$ nm, $c/a=0.668$, and $u=0.304$ are in good agreement with experimental values of 0.301 nm, 0.67, and 0.304 \cite{Bortz:99}. The band structure calculated for special $k$-points is shown in Figure \ref{band}, which is in good agreement with the previous GGA calculations for the host \cite{Park:09}.  It is noteworthy that the band gap is indirect (3.71 eV), and that the relatively steep dispersion near the conduction band minimum (CBM) and valence band maximum (VBM) suggests high mobility of excited carriers as well as a long lifetime.

\section{Result}\label{Rslt}
\begin{figure}[t]
\begin{center}
\includegraphics[width=0.9\linewidth]{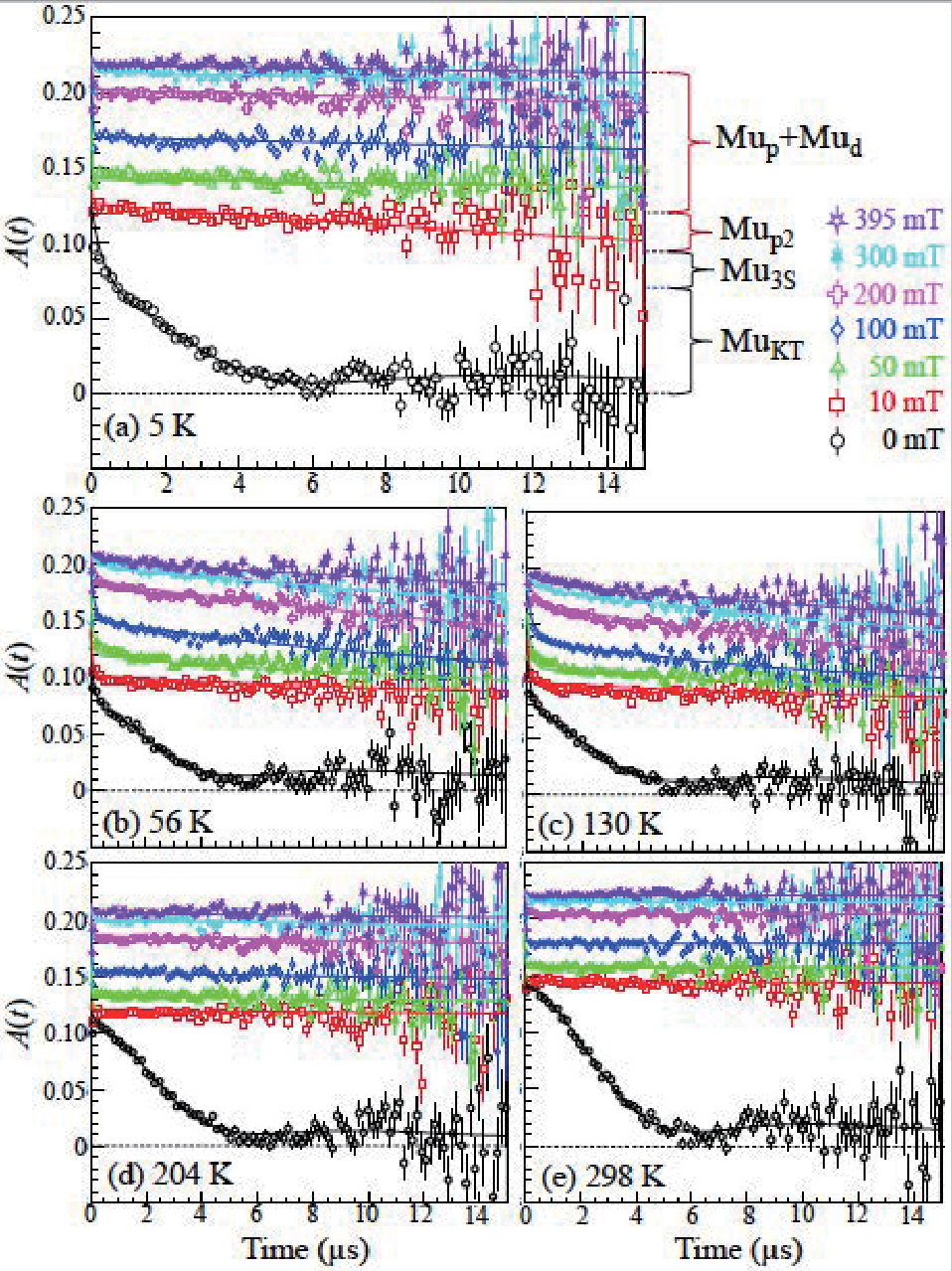}
\caption{$\mu$SR time spectra under zero and longitudinal magnetic fields in MgH$_2$ at typical temperatures. (a) 5 K, (b) 56 K, (c) 130 K, (d) 204 K, (e) 298 K. (a) Right: time spectrum shows an initial paramagnetic Mu (Mu$_{\rm p}$, whose asymmetry is recovered by a longitudinal field) that exhibits conversion to a diamagnetic Mu state (Mu$_{\rm d}$). Another paramagnetic state (Mu$_{\rm p2}$, showing fast relaxation) and a diamagnetic Mu component showing the lineshape well represented by the Kubo-Toyabe relaxation function (Mu$_{\rm KT}$) can be seen.  The solid line shows the result of the global fit based on the model (see text).}
\label{tspec}
\end{center}
\end{figure}

Figure \ref{tspec} shows the observed ZF/LF-$\mu$SR time spectra at characteristic temperatures. It can be seen that $A(t)$ mainly consists of two components, a paramagnetic component (labeled Mu$_{\rm p}$) whose asymmetry is gradually recovered by LF $\simeq0.05$--0.4 T and a component described by the Kubo-Toyabe function at ZF [equation (\ref{gkt})] \cite{Hayano:79}.  At lower temperatures, two more components emerges in the low LF range [Figure \ref{tspec}(a), labeled Mu$_{\rm p2}$ and Mu$_{\rm 3S}$, respectively], suggesting that the spectra consist of four components in total. 

It is also noticeable in Figure \ref{tspec}(a) that these four components appear sequentially with increasing LF. This indicates that their hyperfine interactions (including those with nuclear magnetic moments) are mutually distinct, and each component can be identified by the magnitude of its characteristic LF.  Therefore, even though these spectra do not show a significant change with temperature, a reliable analysis can be performed by building a model that reproduces their detailed LF dependence.   In fact, it turns out from the LF-dependence that Mu$_{\rm p}$ and Mu$_{\rm p2}$ components are paramagnetic states, while Mu$_{\rm 3S}$ and Mu$_{\rm KT}$ are diamagnetic states. The Mu$_{\rm 3S}$ component, which shows relatively fast exponential relaxation, is interpreted to be due to the H-Mu-H complex according to the previous report \cite{Umegaki:14}. 

A closer look at the time spectra shows that, in order to correctly reproduce their time evolution, it is necessary to take into account the effects of the various spin/charge dynamics to which these states are subjected. More specifically, the sharp decrease of the initial asymmetry [$A(0)$] in the spectra at 0--10 mT from 0.22 (=$A_{\rm tot}$) to the value roughly corresponding to that of Mu$_{\rm KT}$ [$A_{\rm KT}\sim$0.09--0.14] within $\sim$0.5 $\mu$s at all temperatures suggests that the asymmetry corresponding to the spin-triplet state of Mu$_{\rm p}$ [$\simeq(A_{\rm tot}-A_{\rm KT})/2$, see equation (\ref{MuPzr}) in Appendix A] is lost due to the fast spin/charge exchange reaction with the excited carriers \cite{Patterson:88} or some epithermal processes \cite{Walker:83,Percival:85,Storchak:99}. In Figures \ref{tspec}(a), (d) and (e), there are non-relaxing asymptotic components in the spectra with LF $\ge50$ mT, whose asymmetry recovers with increasing LF, suggesting a transition to the diamagnetic state (labeled Mu$_{\rm d}$) via an irreversible charge conversion reaction \cite{Kadono:03}.  Furthermore, in the spectra in Figures \ref{tspec}(b) and (c), the final Mu$_{\rm d}$ state shows a slow relaxation where the decay rate is almost independent of LF, suggesting that fast fluctuations of the hyperfine field occur with increasing temperature even in the final state.

\begin{figure}[t]
\begin{center}
 \includegraphics[width=0.85\linewidth]{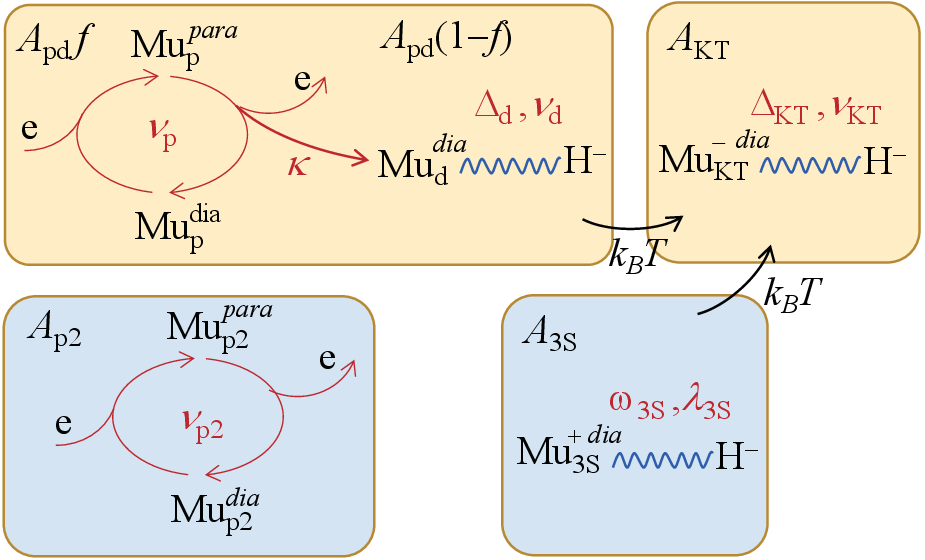}
\caption{A schematic diagram of the phenomenological model involving Mu$_{\rm p}$/Mu$_{\rm d}$, Mu$_{\rm KT}$, Mu$_{\rm p2}$, and Mu$_{\rm 3S}$ states and their spin/charge exchange reactions represented by equations (\ref{gz})--(\ref{gd}). Here, the paramagnetic state are expressed as {\it para} and the diamagnetic states as {\it dia}.  Since all the initial Mu states are relaxed-excited states (non-equilibrium), the arrow labeled ``$kT$" represents the process of approaching thermal equilibrium by annealing as the temperature is increased.}
\label{model}
\end{center}
\end{figure}

Based on the above observations, the time spectra were analyzed by the least-squares curve fits using the following phenomenological model for the spin/charge dynamics.
\begin{eqnarray}
A(t)&=&A_{\rm pd} G_{\rm pd}(t)+A_{\rm p2} G_{\rm p2}(t) \nonumber\\
& & +A_{\rm KT} G_z^{\rm KT}(t)+A_{\rm 3S}G_z^{\rm 3S}(t),\label{gz}
\end{eqnarray}
\begin{equation}
G_{\rm pd}(t)= fe^{-(\lambda_{\rm p}+\kappa)t}g_z(x_{\rm p})+(1-f) e^{-\lambda_{\rm d} t},
\end{equation}
\begin{eqnarray}
\lambda_{\rm p}=\frac{\nu_{\rm p}}{2(1+x_{\rm p}^2)}\:,
f=\frac{\lambda_{\rm p}}{\lambda_{\rm p}+\kappa}\:,\:\:
\lambda_{\rm d}\simeq\frac{2\Delta_{\rm d}^2\nu_{\rm d}}{\omega^2+\nu_{\rm d}^2}\:,\label{gp}
\end{eqnarray}
\begin{equation}
G_{\rm p2}(t) =  e^{-\lambda_{\rm p2}t}g_z(x_{\rm p2}),\:\:\:
\lambda_{\rm p2}\simeq\frac{\nu_{\rm p2}}{2(1+x_{\rm p2}^2)}\:,\label{gd}
\end{equation}
where $A_{\rm pd}$ is the sum of partial asymmetry for Mu$_{\rm p}$ and Mu$_{\rm d}$ components, $A_{\rm p2}$, $A_{\rm KT}$, and $A_{\rm 3S}$ are the partial asymmetry of the respective Mu$_{\rm p2}$, Mu$_{\rm KT}$, and  Mu$_{\rm 3S}$ ($A_{\rm tot}=A_{\rm pd}+A_{\rm p2}+A_{\rm KT}+A_{\rm 3S}$), $G^{\rm KT}_z(t;\Delta_{\rm KT},\nu_{\rm KT})$ is the Kubo-Toyabe function for Mu$_{\rm KT}$ with $\Delta_{\rm KT}$ and $\nu_{\rm KT}$ being the linewidth and fluctuation frequency, $G_z^{\rm 3S}(t;\omega_{\rm 3S},\lambda_{\rm 3S})$ is the relaxation function for the H-Mu-H complex state, $f$ is the fractional yield of Mu$_{\rm p}$, $\nu_{\rm p}$ is the spin/charge exchange rate, $\kappa$ is the charge conversion reaction rate from Mu$_{\rm p}$ to Mu$_{\rm d}$ \cite{Kadono:03}, $g_z(x)= (\frac{1}{2}+x^2)/(1+x^2)$ is the initial polarization of the paramagnetic Mu, $x_{\rm p}=B_0(\gamma_\mu-\gamma_e)/\omega_{\rm p}$, $\omega_{\rm p}$ is the hyperfine parameter (angular frequency) of Mu$_{\rm p}$, $\Delta_{\rm d}$ is the nuclear magnetic dipolar linewidth, $\omega =\gamma_\mu B_0$, and $\nu_{\rm d}$ is the fluctuation rate of $\Delta_{\rm d}$ including the effect of spin/charge exchange reaction.  The parameter $f$ is determined by the competition between $\kappa$ (varies with temperature) and $\lambda_{\rm p}$, and it also depends on LF.  For Mu$_{\rm p2}$, the relaxation due to the spin/charge exchange reaction ($\lambda_{\rm p2}$) is considered as in the case of Mu$_{\rm p}$. $\nu_{\rm p2}$ is the spin/charge exchange rate for the Mu$_{\rm p2}$ component, $x_{\rm p2}=B_0(\gamma_\mu-\gamma_e)/\omega_{\rm p2}$, and $\omega_{\rm p2}$ is the hyperfine parameter of Mu$_{\rm p2}$.  [See Appendix A for details on these relaxation functions, including the derivation of the equations (\ref{gp})--(\ref{gd}).] A schematic representation of the relationship between the four states and the meaning of the parameters in the equations (\ref{gz})--(\ref{gd}) is shown in Figure \ref{model}.

\begin{figure}[t]
\begin{center}
\includegraphics[width=1.0\linewidth]{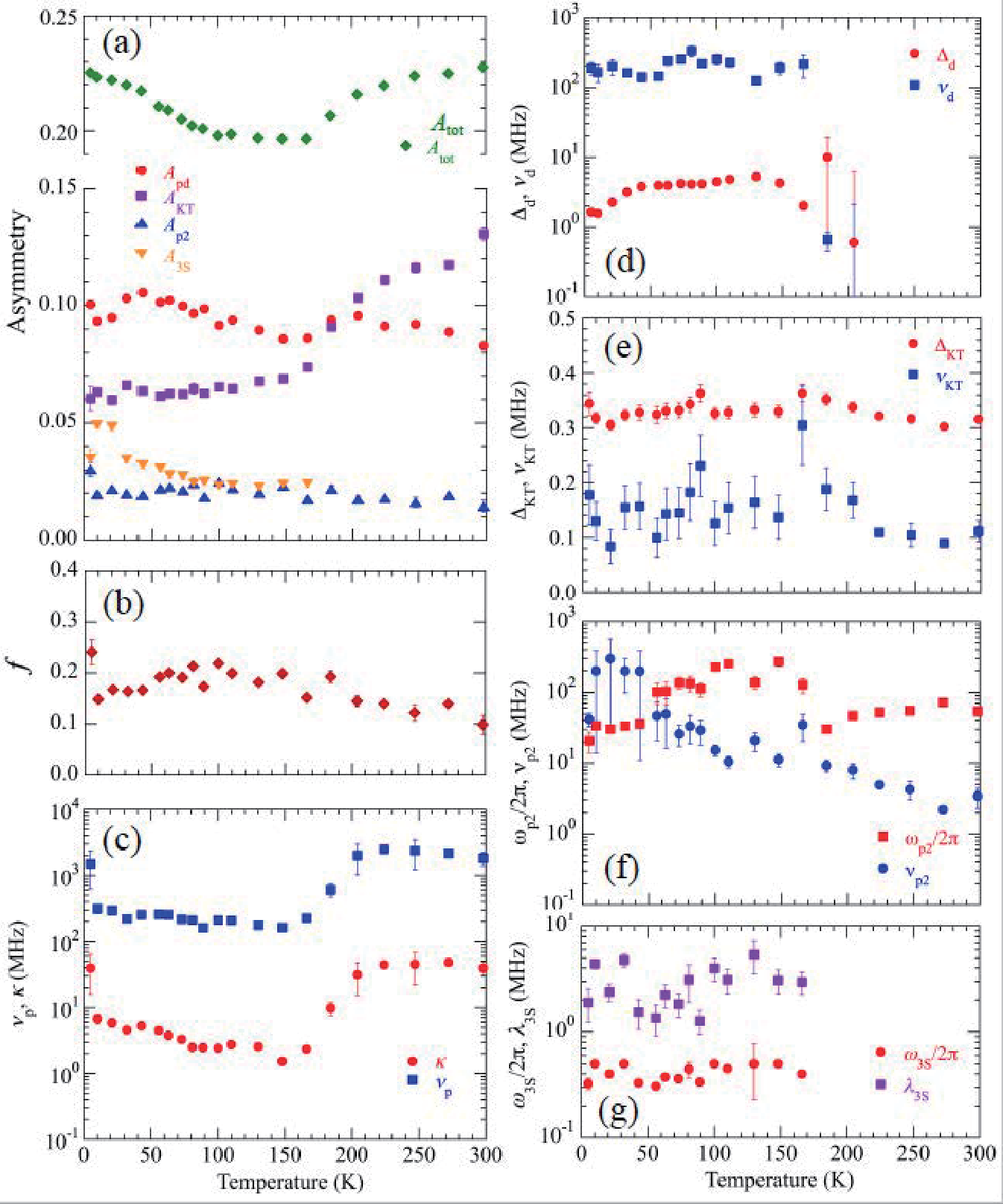}
\caption{Temperature dependence of the parameters obtained from the analysis of ZF/LF-$\mu$SR time spectra: (a) the initial asymmetry of each component, (b) the fractional yield of Mu$_{\rm p}$, (c) spin/charge fluctuation frequency of the Mu$_{\rm p}$ component ($\nu_{\rm p}$) and conversion rate to Mu$_{\rm d}$ ($\kappa$), (d), (e) nuclear dipolar linewidths of Mu$_{\rm d}$ and Mu$_{\rm KT}$ ($\Delta_{\rm d}$ and $\Delta_{\rm KT}$) and their fluctuation frequencies ($\nu_{\rm d}$ and $\nu_{\rm KT}$), (f) hyperfine parameters of the Mu$_{\rm p2}$ component ($\omega_{\rm p2}$) and its spin fluctuation frequency ($\nu_{\rm p2}$), (g) the precession frequency ($\omega_{\rm 3S}$) and relaxation rate ($\lambda_{\rm 3S}$) of the H-Mu-H three-spin system.}
\label{params}
\end{center}
\end{figure}

In the curve-fit analysis, $\omega_{\rm p}$ showed a considerable temperature dependence [$\omega_{\rm p}/2\pi$ ranging from 0.52(4) to 1.73(5) GHz] in correlation with $\nu_{\rm p}$ and $\kappa$ [see Figure~\ref{omgp} in Appendix A]. In particular, in the region where $\nu_{\rm p}\gtrsim\omega_{\rm p}/2\pi$ [above $\sim$150 K, see Figure \ref{params}(c)], $\omega_{\rm p}$ dropped sharply with decreasing accuracy. This is because the effective hyperfine parameter is reduced when the duration for a $1s$ orbital electron to exert a continuous hyperfine field on $\mu^+$ ($\sim1/\nu_{\rm p}$) becomes shorter than the spin rotation period due to the hyperfine interaction ($\sim 2\pi/\omega_{\rm p}$) \cite{Patterson:88,Nosov:65,Chow:00}. Such an effect, also known as ``motional narrowing'' in magnetic resonance, could account for the temperature dependence of $\omega_{\rm p}$ correlated with $\nu_{\rm p}$. In the following, in order to examine the temperature variation of other parameters without the effect of $\omega_{\rm p}$ ambiguity, the analysis was performed with $\omega_{\rm p}/2\pi$ fixed at 0.721 GHz, the average value below 50 K where $A_{\rm tot}$ was also nearly constant. 
 
 Although the result still showed a strong correlation between $\nu_{\rm p}$ and $\kappa$, this can be reasonably understood by assuming that the charge conversion process occurs as part of the spin/charge exchange reaction (Figure \ref{model}, top left).
The Mu$_{\rm 3S}$ signal was hard to be discerned from those of other components above $\sim$150 K, and $A_{\rm 3S}$ was set to zero for the analysis of data at higher temperatures.
Meanwhile, since $\omega_{\rm p2}$ showed a tendency of changing with temperature, we assumed that the hyperfine interaction of the second paramagnetic component Mu$_{\rm p2}$ could change with temperature (see below). As indicated by the solid curves in Figure \ref{tspec}, this phenomenological model reasonably reproduces the experimental data including the LF dependence at all temperatures. The temperature dependence of the parameters obtained from the curve-fits is shown in Figure \ref{params}.

The electronic state of Mu$_{\rm p}^0$ is characterized by a relatively large hyperfine parameter ($\omega_{\rm p}/\omega_{\rm vac}\sim$0.12--0.39).
As is evident in Figures \ref{params}(a), (b) and (c), the decrease in $A_{\rm pd}$ and total signal amplitude $A_{\rm tot}$ seen over 50--150 K is in line with the decrease in conversion rate $\kappa$. On the other hand, $\nu_{\rm p}$ appears to be almost constant over this temperature range. Therefore, the decrease of $A_{\rm pd}$ and $A_{\rm tot}$ suggests that the spin relaxation is induced by the fast spin/charge exchange reaction ($\nu_{\rm p}$) as the lifetime of the Mu$_{\rm p}$ state ($\kappa^{-1}$) becomes longer.  The magnitude of $\Delta_{\rm d}$ provides a clue to the local electronic structure of the final Mu$_{\rm d}$ state. As shown in Figure \ref{params}(d), the average value of $\Delta_{\rm d}$ in the 50--150 K region is 2.71(2) MHz.

Regarding Mu$_{\rm KT}$ characterized by the quasi-static Kubo-Toyabe relaxation, the magnitude of the linewidth $\Delta_{\rm KT}$ again provides a strong clue to the microscopic local structure. As shown in Figure \ref{params}(e), $\Delta_{\rm KT}$ is 0.30--0.36 MHz in all temperature regions, while it is calculated to be greater than 0.63 MHz for any interstitial site without lattice relaxation.  Although an exotic Mu state in which Mu is delocalized over two neighboring sites has been proposed in previous studies \cite{Sugiyama:19}, another possibility is more likely from DFT calculations for H, as discussed below.

While the accuracy of the deduced parameters shown in Figure \ref{params}(f) seems limited due to its small amplitude ($A_{\rm p2}/A_{\rm tot}\simeq$5--10\%), we find for the Mu$_{\rm p2}$ component that the entire time spectrum can be consistently reproduced when we assume that the relaxation is that exhibited by a paramagnetic state with $\omega_{\rm p2}/2\pi=$ 20(7)--274(41) MHz. The magnitude of $\omega_{\rm p2}$ is a few \% of $\omega_{\rm vac}$ and significantly smaller than $\omega_{\rm p}$. It is noticeable that $\omega_{\rm p2}$ shows an increase of almost one order of magnitude in the 50--150 K region from both low and high temperatures.  As discussed before, the effective hyperfine parameter would be reduced when $\nu_{\rm p2}\gtrsim\omega_{\rm p2}/2\pi$, which seems to be the case for the change below $\sim$50 K.  On the other hand, the decrease above 150 K is not accompanied by such an increase in $\nu_{\rm p2}$. Therefore, this may suggest a transition to a new paramagnetic state. It should be noted, however, that artifacts from the analysis with fixed $\omega_{\rm p}$ may have affected the results.

Finally, let us give a brief overview on the Mu$_{\rm 3S}$ state.  The corresponding signal exhibits relatively large relaxation rate at all temperatures, $\lambda_{\rm 3S}\gg\omega_{\rm 3S}$, which makes it difficult to clearly identify the oscillatory signal characteristic to the three-spin systems. While the mean value of $\omega_{\rm 3S}/2\pi$ averaged over temperature can be derived to be 0.41(2) MHz [$\omega_{\rm 3S}=2.6(1)$ MHz, corresponding H-Mu distance $r_{\rm t}=0.098(5)$ nm], the actual systematic uncertainty is presumed to be larger than that determined statistically.  The signal appears to merge with that of the Kubo-Toyabe component above $\sim$150 K, implying the instability of the H-Mu-H state.

\section{Discussion}
\subsection{Predictions based on the ambipolarity model and DFT calculations}
\begin{figure}[t]
\begin{center}
\includegraphics[width=\linewidth]{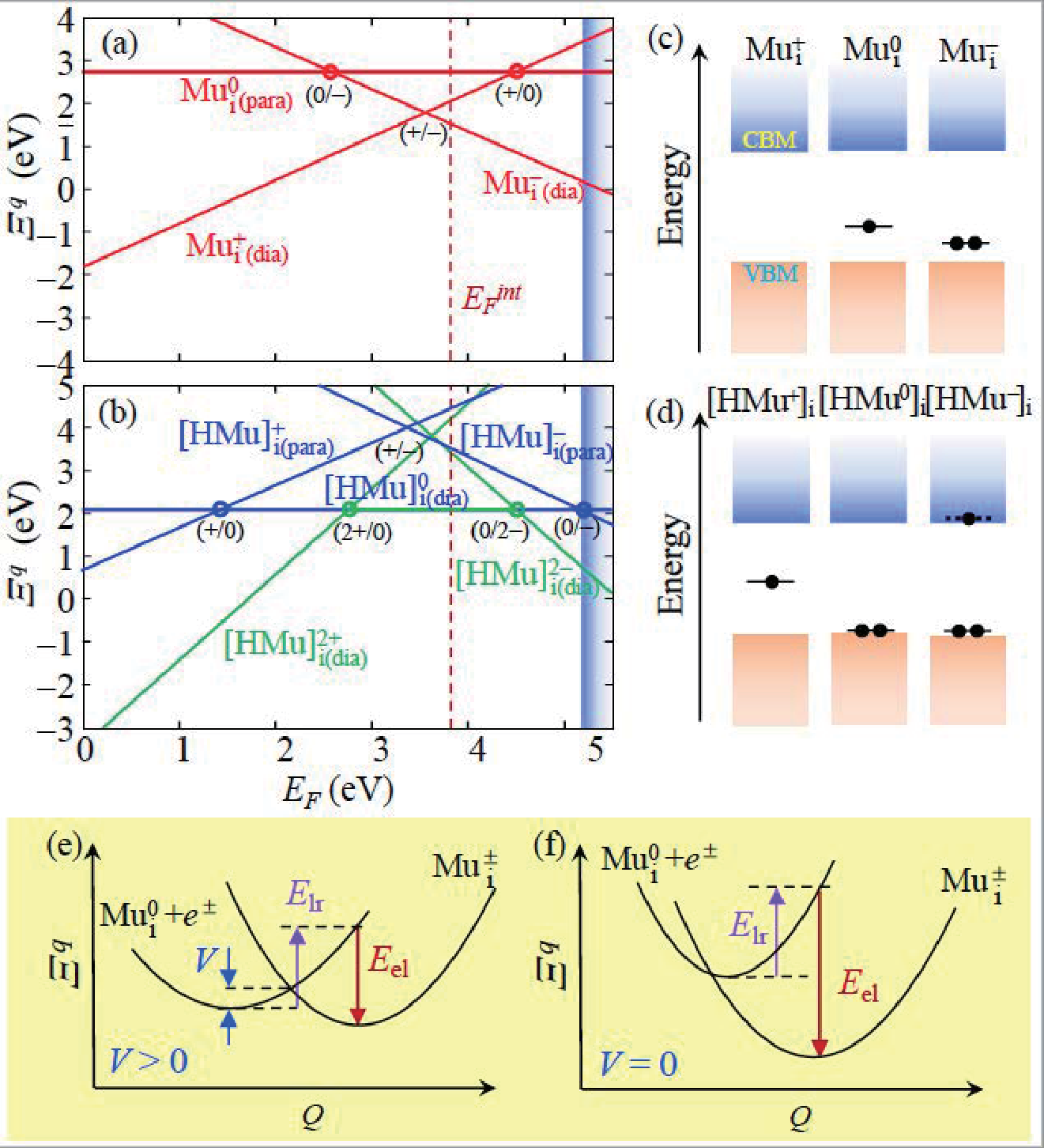}
\caption{The formation energy ($\Xi^q$) of H-related defects vs the Fermi level ($E_F$) in MgH$_2$ obtained by DFT calculations, where (a) is for the case of interstitial H atoms (H$^q_{\rm i}$, $q=0,\pm$), and (b) is for the interstitial molecules ([H$_2]^q_{\rm i}$, $q=0,\pm,2\pm$), respectively corresponding to Mu$_{\rm i}$  and [HMu]$_{\rm i}$. The paramagnetic state are expressed as ``para'' and the diamagnetic states as ``dia''. The levels for acceptor ($E^{0/-}$) and donor ($E^{+/0}$) are determined as cross points between $\Xi^\pm(E_F)$ and $\Xi^0(E_F)$.  The dashed line shows the intrinsic charge neutral level ($E_F^{\rm int}$) obtained from the DFT calculation. (Adapted from Ref.~\cite{Roy:13})  
Schematic band diagrams for the single electron energy associated with the donor/acceptor levels (single-charge conversion levels only) in (a) and (b)  are shown in (c) and (d), respectively.  The level indicated by a dashed line near the CBM in (d) is a shallow state suggested from DFT calculations. For the corresponding local electronic structure, see Figures \ref{Hint},\ref{Hmol}.  (e), (f) Schematic of $\Xi^q$ versus configuration coordinate ($Q$) for Mu$^0_{\rm i}$ and the charged states, where $E_{\rm lr}$ is the lattice relaxation energy, $E_{\rm el}$ is the Coulomb energy, and $V$ is the potential barrier between the two states. Mu$^0_{\rm i}$ is expected to be observed as a relaxed excited state for (e) $V>0$ but not for  (f) $V=0$.}
\label{FE}
\end{center}
\end{figure}

In order to compare these results with the predictions of ambipolarity model plus DFT calculations, we extracted information related to the formation energy of H-related defects from the literature \cite{Roy:13} and summarized it for Mu as shown in Figure \ref{FE}. The thermodynamic defect formation energy $\Xi^q$ of H$^q$ (with $q=0,\pm$ being the valence) in a certain material is evaluated by DFT calculations using the following formula,
\begin{equation}
\Xi^q(E_F)=E_{\rm tot}[{\rm H}^q]-E_{\rm tot}[{\rm host}]+qE_F+n_\mathrm{H}\mu_\mathrm{H},\label{Hfm}
\end{equation}
where $E_{\rm tot}[{\rm H}^q]$ and $E_{\rm tot}[{\rm host}]$ are the total energies calculated for the supercell with and without H$^q$ between lattices, respectively, and $E_F$ is the Fermi energy. 
$\mu_{\rm H}$ is the H chemical potential that represents the change in $\Xi^q$ when H is removed ($n_{\rm H}=-1$) or added ($n_{\rm H}=+1$) to the host crystal upon defect formation. We assume that Mu's experimental results correspond to calculations in the limit where the added H is dilute, i.e., under H-poor conditions in terms of chemical potential. This is supported by the fact that the sample was immediately transferred from the inert gas-tight container to a vacuum environment, where the H partial pressure in the sample environment was kept negligibly small. In addition, previous $\mu$SR experiments have shown that the yield of H-Mu-H complexes, which assumes the pre-existence of extra H$_{\rm i}$ atoms, is small for as-prepared samples and increases with dehydrogenation processes such as milling and oxidation catalyst incorporation \cite{Umegaki:14}.  The present experimental result is similar to that of this as-prepared sample (see below), suggesting that the amount of extra H was small.  We also point out that a part of the extra H$_{\rm i}$ contributing to the H-Mu-H complexes could be that ejected from the lattice point by elastic scattering with incident muons (see Appendix B).

Based on Figure \ref{FE} and the ambipolarity model, the following predictions can be made.
Firstly, regarding the general trend of $\Xi^q$ obtained for interstitial H (H$_{\rm i}$) shown in Figure \ref{FE}(a), the $E^{0/-}$ and $E^{+/0}$ levels are on the higher energy side than the $E^{+/-}$ level, indicating the negative $U$ character due to strong electron-phonon interactions \cite{Anderson:75}. Moreover, both $E^{0/-}$ and $E^{+/0}$ form apparently deep ingap levels, in which the existence of two paramagnetic Mu states (Mu$^0_{\rm i}$'s) may be expected. However, as shown in Figures \ref{FE}(e) and (f), whether Mu$^0_{\rm i}$ can actually exist as a relaxed excited state depends on the presence of a potential barrier $V$ at the transition to the charged states. As we will see below,  $V>0$ is inferred for the transition to Mu$^+_{\rm i}$, while $V\simeq0$ is suggested for the transition to Mu$^-_{\rm i}$ because the Mu$^0_{\rm i}$ state corresponding to the acceptor level is not observed in the experiment (although this does not rule out the possibility of short-lived Mu$^0_{\rm i}$).  Note that the one-electron energy levels shown in Figure \ref{FE}(c) correspond to $E_{\rm el}$, not $E^{-/0}$. Regarding the H$_{\rm i}^-$ state (corresponding to Mu$_{\rm i}^-$), it is predicted to accompany significant local lattice relaxation; H$_{\rm i}^-$ pushes away the surrounding ligand H$^-$ atoms by 0.052 nm while it attracts Mg$^{2+}$ \cite{Park:09}.  On the other hand, the charged donor-like H (H$_{\rm i}^+$, corresponding to Mu$_{\rm i}^+$) binds strongly to the ligand H, forming a neutral HMu molecule corresponding to the onsite H$_2$ molecule \cite{Park:09}.  The single-electron energetics are shown in Figure \ref{FE}(c).

\begin{figure}[t]
\begin{center}
\includegraphics[width=0.9\linewidth]{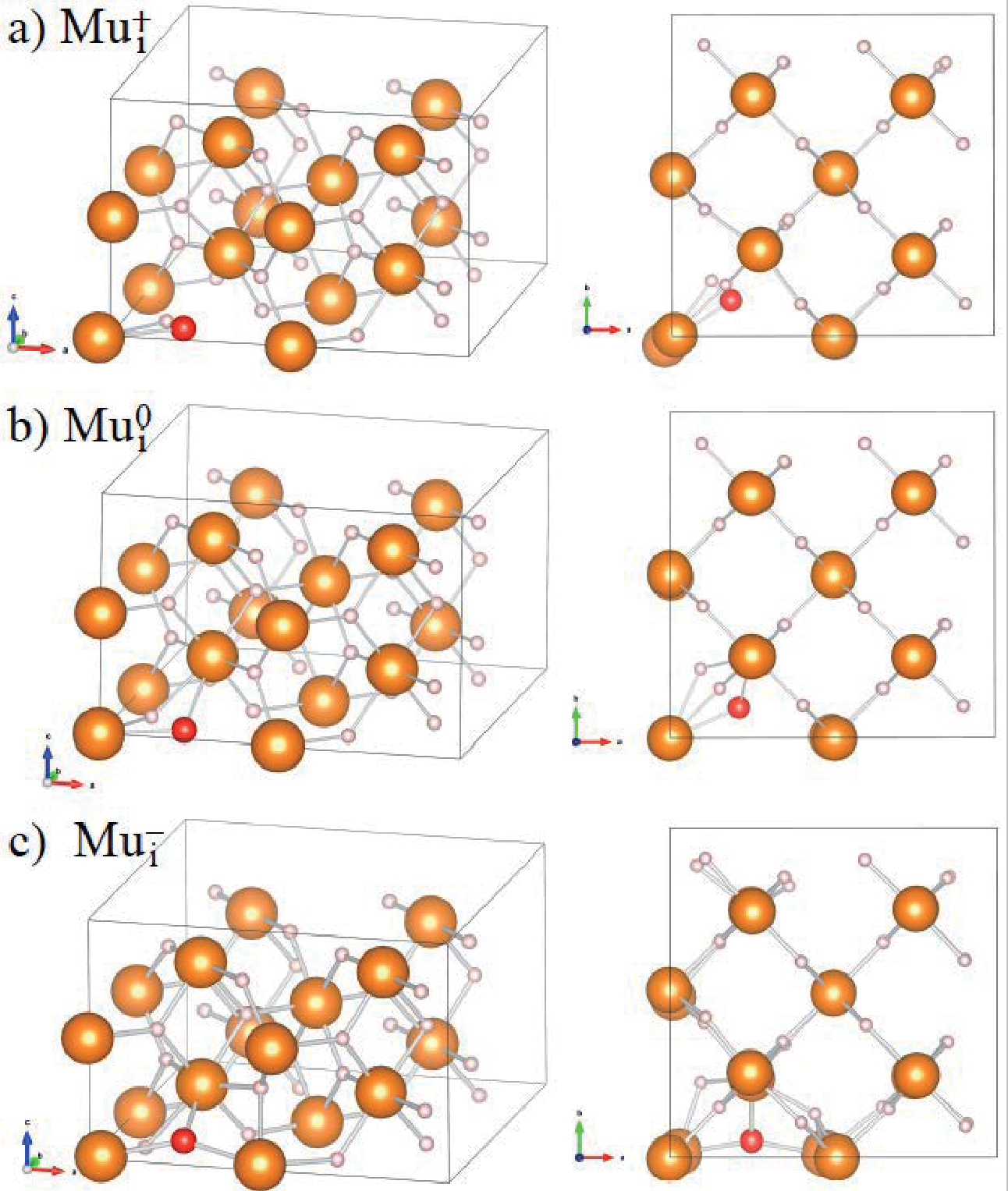}
\caption{Local structure of interstitial H (H$_{\rm i}$) in (a) positive, (b) neutral, and (c) hydride states which are substituted with Mu (red balls), viewed from $b$ (left) and $c$ (right) axes, respectively. }
\label{Hint}
\end{center}
\end{figure}

Secondly, provided that Mu can form an interstitial molecule with a pre-existing H$_{\rm i}$ atom (denoted as [HMu]$_{\rm i}$), the electronic states for the interstitial H$_2$ molecules ([H$_2]_{\rm i}$) predicted by the DFT calculations can provide the basis for discussing those of [HMu]$_{\rm i}$.  As shown in Figure \ref{FE}(b), the [HMu]$_{\rm i}^0$ state is stable over a wide $E_F$ region in the band gap, showing lower formation energy than that for [HMu]$_{\rm i}^+$ and [HMu]$_{\rm i}^-$ (i.e., positive-$U$), and can exist even in thermal equilibrium. The $E^{0/-}$ and $E^{+/0}$ levels are closer to the band edge than for the isolated Mu, and $E^{0/-}$ appears to overlap the conduction band minimum (CBM). Regarding the local structure, the [HMu]$_{\rm i}^+$ molecule is pulled to ligand H$^-$ to form a paramagnetic complex state (see below). The double charge conversion levels for [HMu]$_{\rm i}^{2+}$ and [HMu]$_{\rm i}^{2-}$ are closer to the charge neutral level ($E_F^{\rm int}$) and form deeper levels. The  [HMu]$_{\rm i}^{2+}$ molecular states are unstable, and both H and Mu are presumed to form onsite neutral H$_2$ and HMu molecules with ligand H$^-$, respectively. The corresponding single-electron energetics are shown in Figure \ref{FE}(d).

\subsection{Local electronic structure of each Mu state}
\begin{figure}[t]
\begin{center}
 \includegraphics[width=0.95\linewidth]{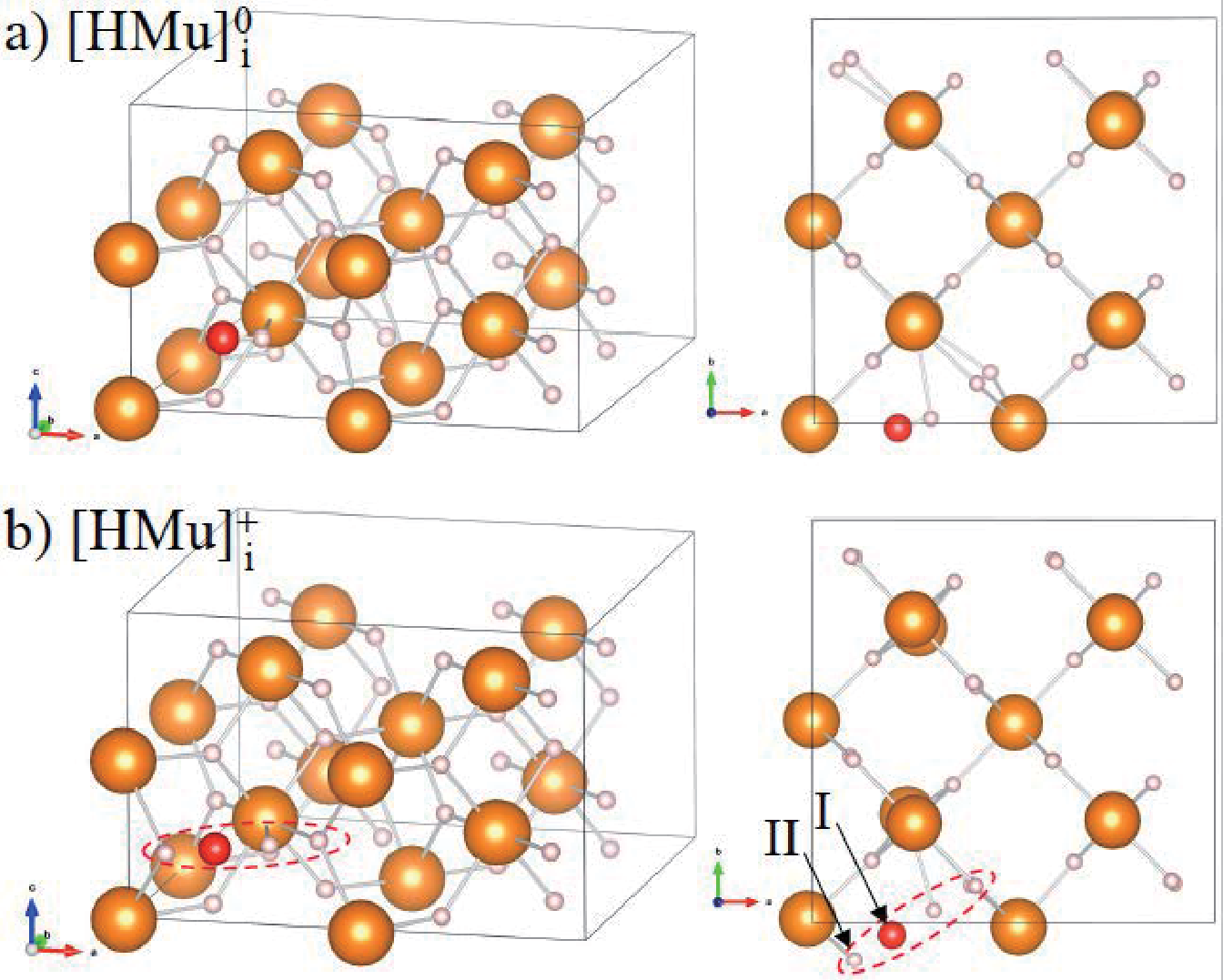}
\caption{Local structure of an interstitial hydrogen molecule in (a) [H$_2]_{\rm i}^0$ and (b) [H$_2]_{\rm i}^+$ states with one H atom substituted by Mu (red balls), viewed from $b$ (left) and $c$ (right) axes, respectively.  There are two inequivalent H denoted by labels I and II for (b), where the case with H$_{\rm I}$ replaced with Mu is shown. The hyperfine parameters for (b) are distributed over four H atoms indicated by a dashed ellipse. }
\label{Hmol}
\end{center}
\end{figure}

In order to obtain further clues about the correspondence between these DFT predictions and the experimental results for Mu, we have investigated the local defect structure associated with these acceptor/donor levels in each of the above two cases by QE calculations. The obtained local structures for H$_{\rm i}$ and [H$_2]_{\rm i}$ are shown in Figures \ref{Hint} and \ref{Hmol}, respectively, where those for H$_{\rm i}$ are in good agreement with earlier GGA results using VASP code \cite{Park:09}.  We found that the $E^{+/0}$ level associated with H$_{\rm i}$ (= Mu$_{\rm i}$) and that of the [H$_2]_{\rm i}$ molecule (= [HMu]$_{\rm i}$) have a finite unpaired electron density in the band gap (i.e., the paramagnetic states). This is consistent with the experimental observation of two paramagnetic states, Mu$_{\rm p}$ and Mu$_{\rm p2}$. The hyperfine parameters calculated by GIPAW module of the QE suite for these states are listed in Table \ref{hfp}. They exhibit a common character that the Fermi contact interaction is predominant.  For the other diamagnetic states, the nuclear dipolar width $\Delta$  are summarized in Table \ref{delta_c}. 

Given that the Mu states associated with the defect levels in Figure \ref{FE}(c) and (d) are meaningful only in the non-equilibrium state, there is no guarantee that the electronic structures of Mu discussed below correspond to those in thermal equilibrium. However, since the latter are generally not significantly altered by $E_F$, it is reasonable to assume that, among the relaxed-excited states, those that can also exist in thermal equilibrium have the same electronic structure as the latter. Specifically, for the Mu$_{\rm i}$ states (negative-$U$ centers), Mu$^0_{\rm i}$  has no corresponding state in thermal equilibrium, but Mu$^\pm_{\rm i}$ states correspond to those in thermal equilibrium (and to H$^\pm_{\rm i}$); for the [HMu]$_{\rm i}$ molecules (positive-$U$ centers), all states possibly correspond to those in thermal equilibrium (and to [H$_2]_{\rm i}$). In light of the above considerations,  we discuss the correspondence between these levels and the experimentally observed Mu states.

\begin{table}[b]
\caption{Calculated hyperfine parameters for the paramagnetic states: H$^0_{\rm i}$ [corresponding to Figure \ref{Hint}(b)] and those for each H in [H$_2]_{\rm i}^+$ molecule [I and II in Figure \ref{Hmol}(b)]. The values for Mu must be multiplied by 3.18. The bottom rows show experimental values for paramagnetic Mu states in comparison with those for candidate paramagnetic states inferred from DFT calculations.}\label{hfp}
  \renewcommand\arraystretch{1.1}
\begin{center}
\footnotesize
\lineup
\begin{tabular}{@{}c|c|ccc}
\br
\multicolumn{2}{c|}{\ }  & \multicolumn{2}{c}{Magnetic dipolar term (MHz)} & Principal axis\\
\cline{3-5}
\multicolumn{2}{c|}{\ }& $\omega_{xx}/2\pi$ &  \-0.0295    & $(0,  0,  1)$\\
\multicolumn{2}{c|}{H$_{\rm i}^0$} &  $\omega_{yy}/2\pi$ &  \-1.8829  & $(\m0.5514, -0.8342,  0)$\\
\multicolumn{2}{c|}{\ } &  $\omega_{zz}/2\pi$ &   $1.9125$    & $(-0.8342, -0.5514,  0)$\\
 \cline{3-5}
\multicolumn{2}{c|}{\ }  & \multicolumn{2}{c}{Fermi contact term (MHz)} & \\
 \cline{3-5}
\multicolumn{2}{c|}{\ }  &  $ \omega_c/2\pi$ & $630.013$ & ----\\
\mr
 \multirow{11}{*}{[H$_2]_i^+$}& & \multicolumn{2}{c}{Magnetic dipolar term (MHz)} & Principal axis\\
 \cline{3-5}
&  \multirow{4}{*}{I} & $\omega_{xx}/2\pi$ &  \-19.1146  & $(\m0.4991, -0.8665,  0)$\\
& & $\omega_{yy}/2\pi$ &  \-19.5940  & $(0, 0, 1)$\\
& & $\omega_{zz}/2\pi$ &   38.7086    & $(-0.8665, -0.4991, 0)$\\
\cline{3-5}
& & \multicolumn{2}{c}{Fermi contact term (MHz)} & \\
\cline{3-5}
& & $ \omega_c/2\pi$ & $267.150$ & ----\\
\cline{2-5}
& & \multicolumn{2}{c}{Magnetic dipolar term (MHz)} & Principal axis\\
\cline{3-5} 
& \multirow{4}{*}{II} &  $\omega_{xx}/2\pi$ &  \-3.1894   & $(0,  0,  1)$\\
& &  $\omega_{yy}/2\pi$ &  \-4.7335  & $(\m0.3198, -0.9475,  0)$\\
& &  $\omega_{zz}/2\pi$ &   1.9125    & $(-0.9475, -0.3198,  0)$\\
\cline{3-5}
& & \multicolumn{2}{c}{Fermi contact term (MHz)} & \\
\cline{3-5}
& &   $ \omega_c/2\pi$ & $271.750$ & ----\\
\br
\multicolumn{2}{c|}{Mu}& \multicolumn{2}{c}{Mu$^0_{\rm i}$ ($\omega_c$)} & [HMu]$^+_{\rm i}$ ($\omega_c$)\\
\multicolumn{2}{c|}{h.f. (GHz)} &  \multicolumn{2}{c}{2.003} & (I) 0.849 / (II) 0.864 \\
\mr
\multicolumn{2}{c|}{\ }& \multicolumn{2}{c}{Mu$_{\rm p}$ ($\omega_{\rm p}$)} &  Mu$_{\rm p2}$  ($\omega_{\rm p2}$)\\
\multicolumn{2}{c|}{Exp. (GHz)}&  \multicolumn{2}{c}{0.52(4)--1.73(5)} & 0.020(7)--0.274(41)\\
\br
\end{tabular}
 \end{center}
\end{table}

\begin{table}[b]
\caption{Calculated nuclear dipolar linewidth $\Delta$ for the diamagnetic Mu states. The bottom row shows experimental values for diamagnetic Mu corresponding to each state in the upper row.}\label{delta_c}
\begin{center}
\footnotesize
\lineup
\begin{tabular}{cccc}
\br
 & \multicolumn{2}{c}{Mu$_{\rm i}$ } & [HMu]$_{\rm i}$\\
\hline
 Electronic state & Mu$_{\rm i}^+$ & Mu$_{\rm i}^-$ & [HMu]$^{0,2\pm}_{\rm i}$\\
 $\Delta$ (MHz) & 3.64 & 0.3685 & 3.2--3.9 \\
\mr
   \multirow{2}{*}{Exp. (MHz)}  & Mu$_{\rm d}$ ($\Delta_{\rm d}$) & Mu$_{\rm KT}$ ($\Delta_{\rm KT}$) & Mu$_{\rm 3S}$ ($\omega_{\rm 3S}$)\\
 &  2.71(2) &  0.330(4) & 2.6(1)\\
 \br
 \end{tabular}
 \end{center}
\end{table}

\vspace{5pt}
\noindent
$\bullet\:\:{\bf Mu}_{\bf p}$ ${\bf and}$ ${\bf Mu}_{\bf d}$\\
\indent As mentioned earlier, there is considerable uncertainty in the experimental estimate of $\omega_{\rm p}$ (see the bottom row of Table \ref{hfp} and Figure \ref{omgp}), making it difficult to determine whether Mu$_{\rm p}$ corresponds to Mu$^0_{\rm i}$ or [HMu]$^+_{\rm i}$ from the value of $\omega_{\rm p}$ alone.  Since Mu$_{\rm p}$/Mu$_{\rm d}$ is considered to be a pair with Mu$_{\rm KT}$ from the ambipolarity model, let us first discuss the attribution of Mu$_{\rm d}$.  The empirical rule in the model suggests that the yields of paired acceptor-like and donor-like Mu are nearly equal to satisfy overall charge neutrality.  Given that $A_{\rm pd}$ is comparable with $A_{\rm KT}$ over the wide temperature range and that  Mu$_{\rm KT}$ is attributed to Mu$_{\rm i}^-$ (see below), it is reasonable to ascribe Mu$_{\rm d}$ to Mu$_{\rm i}^+$, and thus Mu$_{\rm p}$ is ascribed to Mu$_{\rm i}^0$.  The latter is also in line with the fact that the maximum value of $\omega_{\rm p}$ [$=2\pi\times$1.73(5) GHz, realized when the ``motional narrowing'' effect is negligible] is about 86\% of the value predicted from $\omega_c$ of H$^0_{\rm i}$ ($=2\pi\times$2.003 GHz), which is closer than that of [H$_2]^+_{\rm i}$ ($=2\pi\times$0.85-0.86 GHz). Furthermore, this result indicates that the potential barrier for the transition from Mu$^0_{\rm i}$ to Mu$^+_{\rm i}$ is positive [$V>0$, see Figure \ref{FE}(e)].

\indent For the experimental value of $\Delta_{\rm d}$, it can be explained when the H-Mu distance is $r=0.083$ nm, about 11\% larger than the p-p distance of 0.078 nm estimated from the QE calculation. The relatively large $\nu_{\rm d}$ (150--230 MHz) suggests that even after Mu$_{\rm p}$ (= Mu$_{\rm i}^0$) destabilizes and transitions to Mu$_{\rm d}$ (= Mu$_{\rm i}^+$), a fast spin/charge exchange reaction (Mu$_{\rm i}^+ +e^- \rightleftarrows$ Mu$_{\rm i}^0$) persists with the excitonic carriers.

\vspace{5pt}
\noindent
$\bullet\:\:{\bf Mu}_{\bf KT}$\\
\indent The linewidth $\Delta_{\rm KT}$ of the Mu$_{\rm KT}$ component [= 0.330(4) MHz on average, see Figure \ref{params}(e)] is in good agreement with 0.3685 MHz expected for Mu$_{\rm i}^-$ that accompanies significant outward relaxation of the surrounding H$^-$.  Considering that the incident $\mu^+$ is accompanied by $\sim$10$^2$ electron-hole pairs (see Appendix B), it may be allowed to assume that the probability of capturing two electrons and taking the anion state is sufficiently high.  As discussed earlier, the reason why the relaxed excited state Mu$^0$ associated with this state is not observed is presumably because there is no potential barrier for the transition  to Mu$_{\rm i}^-$ [$V\simeq0$, see Figure \ref{FE}(f)]. This unusual situation may be related to the large lattice relaxation around Mu$_{\rm i}^-$ and the strong binding energy due to the associated Coulomb interaction.
 Thus, Mu$_{\rm KT}$ is ascribed to the diamagnetic state associated with the effective acceptor level of isolated Mu$_{\rm i}$ (i.e., Mu$_{\rm KT}^- =$ Mu$_{\rm i}^-$).  From the fact that $A_{\rm pd}$ and $A_{\rm KT}$ are comparable and account for the majority of the $\mu$SR signal, we conclude that the major part of the observed relaxed-excited Mu states can be explained by the pair Mu$_{\rm i}^+$ and Mu$_{\rm i}^-$.

\vspace{5pt}
\noindent
$\bullet\:\:{\bf Mu}_{\bf p2}$\\
\indent Apart from the cases below 50 K, where $\nu_{\rm p2}>\omega_{\rm p2}$ and thus $\omega_{\rm p2}$ is not well defined, $\omega_{\rm p2/}2\pi$ is 0.1--0.3 GHz for 50--150 K and 20--50 MHz above $\sim$150 K, which do not correspond to the hyperfine parameters ($\omega_c$) of any state in Table \ref{hfp}. However, given that Mu$_{\rm p}$ corresponds to Mu$^0_{\rm i}$,  the only remaining possibility for Mu$_{\rm p2}$ is to be attributed to [HMu]$^+_{\rm i}$. 

As shown in Figure \ref{Hmol}(b) and Table \ref{hfp}, our QE calculations indicate that the local structure of [H$_2]_{\rm i}^+$ is complicated, resulting in the two inequivalent H sites with different hyperfine parameters (labeled I and II). Since the Fermi contact term may be strongly dependent on the choice of PAW dataset, one cannot rule out the possibility that the calculated value for such a complex local structure may be highly indeterminate. Meanwhile, $\omega_{\rm p2}$ is comparable with the value expected from the dipolar terms: e.g., for [H$_2]^+_{\rm i}$-I, the powder average $[(\omega^2_{xx}+\omega^2_{yy}+\omega^2_{zz})/3]^{1/2}\times 3.18= 87.4$ MHz. Although this may just be a coincidence, it reminds us that such polaronic states have been reported in some wide-gap oxides (e.g., SrTiO$_3$ \cite{Ito:19}, which are supposedly induced by Mu-exciton interaction \cite{Hiraishi:22}), and Mu$_{\rm p2}$ may correspond to such a state.

\vspace{5pt}
\noindent
$\bullet\:\:{\bf Mu}_{\bf 3S}$\\
\indent The ambipolarity model suggests that the Mu$_{\rm 3S}$ state (which is paired with Mu$_{\rm p2}$) corresponds to the diamagnetic [HMu]$_{\rm i}^0$ state associated with the $E^{-/0}$ level near the CBM. As shown in Table \ref{delta_c}, the fact that the value of $\omega_{\rm 3S}$ is comparable with the calculated $\Delta$ also supports this conjecture.  Although the fast relaxation of the signal makes it difficult to examine the local electronic structure in detail, it is plausible that the Mu$_{\rm 3S}$ state consists of the [HMu]$_{\rm i}^0$ molecule bonded to a ligand H$^-$ via the hydrogen bonding. 
 It is likely that the H-Mu-H signal reported earlier also stems from the same complex state. 
As noted in the discussion about the chemical potential of H in equation (\ref{Hfm}), since extra H$_{\rm i}$ must exist beforehand for Mu$_{\rm 3S}$ to be observed, the amplitude of the signal should be proportional to the amount of such H$_{\rm i}$. Therefore, the present experimental result suggests that a relatively small amount of H$_{\rm i}$ existed in the as-prepared sample.

\vspace{5pt}
\noindent
$\bullet\:\:{\bf Dynamical}\:\:{\bf model}$

\begin{figure}[t]
\begin{center}
\includegraphics[width=0.8\linewidth]{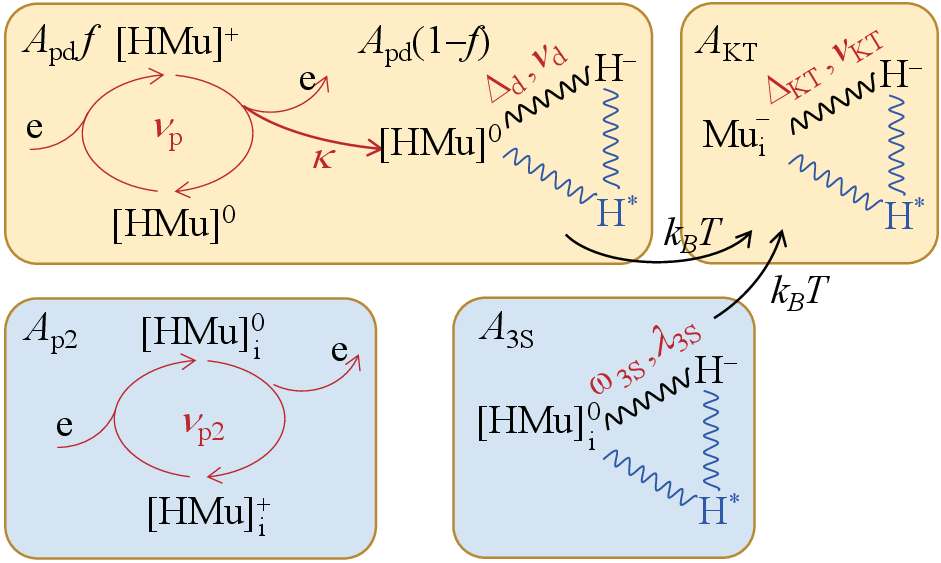}
\caption{The electronic states of Mu estimated on the basis of the ambipolarity and phenomenological model shown in Figure \ref{model}, where Mu$^0_{\rm i}$ = Mu$_{\rm p}$,  Mu$^+_{\rm i}$ = Mu$_{\rm d}$, Mu$_{\rm i}^-$ = Mu$_{\rm KT}^-$, [HMu]$_{\rm i}^+$ = Mu$_{\rm p2}$, and [HMu]$_{\rm i}^0$-H$^-$ = Mu$_{\rm 3S}^+$.   The possibility is shown that Mu$^+_{\rm i}$, Mu$_{\rm i}^-$, and [HMu]$_{\rm i}^0$ states may be affected by spin fluctuations of the nearby H centers in the relaxed-excited state (denoted as H$^*$), a factor not explicitly taken into account in the analysis.}
\label{model2}
\end{center}
\end{figure}

Applying these attributed electronic states of Mu to the phenomenological model, Figure \ref{model} is updated to Figure \ref{model2}. Based on this, we consider a scenario that explains the rough temperature variation of each state observed in Figure \ref{params}. $A_{\rm KT}$ shows a marked increase above 150 K, which is clearly correlated with the increase in the charge transfer reaction rate $\kappa$ from Mu$_{\rm i}^0$ (= Mu$_{\rm p}$) to Mu$_{\rm i}^+$ (= Mu$_{\rm d}$).  This suggests that the transition from Mu$_{\rm i}^0$ to Mu$_{\rm i}^-$  via Mu$_{\rm i}^+$ is accelerated by the annealing effect with increasing temperature. A similar transition from Mu$_{\rm 3S}$ to Mu$_{\rm KT}^-$ is suggested to occur in parallel with that for Mu$_{\rm i}^+$. 
These annealing-induced transitions are explained by the fact that the formation energy $\Xi^q(E_F)$ of Mu$_{\rm i}^-$ is lower than those of Mu$_{\rm i}^0$ around $E_F\simeq E^{+/0}$ and [HMu]$^0_{\rm i}$ around $E_F\simeq E^{0/-}$, as shown in Figure \ref{FE}(a) and (b).

The dynamic model we introduced in equations (\ref{gz})--(\ref{gd}) cannot distinguish whether the fluctuation of the internal magnetic field felt by Mu is due to carrier motion or self-diffusion of Mu. The carrier motion is a strong candidate as the cause of the fluctuations, since the band structure in Figure \ref{band} predicts that both electrons and holes are only weakly localized in MgH$_2$. Regarding the Mu motion, the diamagnetic Mu$_{\rm i}$ is also considered to be localized (at least in the time scale of $\mu$SR) because it is bonded to the ligand H$^-$ (Mu$^+_{\rm i}$) or accompanying large lattice distortions (Mu$^-_{\rm i}$, as inferred from $\nu_{\rm KT}$; see the next section for more detail). On the other hand, it is known that the acceptor-like neutral Mu$^0_{\rm i}$ can exhibit rapid long-range diffusion (e.g., in alkali halides \cite{Kiefl:89,Kadono:90} and III-V semiconductors \cite{Kadono:91a,Kadono:94}).  When Mu$^0_{\rm i}$ exhibits self-diffusion, the fluctuation of the nuclear hyperfine (NHF) interaction between 1$s$ orbital electron and proton nuclear spins also induces Mu$^0_{\rm i}$ spin relaxation. In this case, the magnetic field dependence of the longitudinal relaxation rate is expected to be weaker than that for $\lambda_{\rm p}$ [see equation (\ref{gp})], and to vary significantly with the jump frequency of Mu$^0_{\rm i}$ \cite{Kiefl:89,Kadono:90,Kadono:91a,Kadono:94}. However, $\lambda_{\rm p}$ is so large ($>10^1$ MHz) over the wide temperature range (5--298 K) that such a possibility as the origin of longitudinal relaxation seems unlikely.  Moreover, the present experimental results suggest that even if the Mu$^0_{\rm i}$ state exists, it is converted to Mu$^-_{\rm i}$ in a short lifetime ($\sim\kappa^{-1}$).  Regarding the [HMu]$_{\rm i}$ states, they have a mass one order of magnitude larger than Mu, and is considered to be even more localized than the diamagnetic Mu$_{\rm i}$. Therefore, the values of the dynamical parameters introduced in the model can be interpreted as mainly reflecting the carrier motion.

Finally, regarding the initial electronic states of  Mu$_{\rm p}$/Mu$_{\rm d}$-Mu$_{\rm KT}$ pair (i.e., Mu$_{\rm i}^0$/Mu$_{\rm i}^+$ and Mu$_{\rm i}^-$), neutral and negative ionic states are predominant as a whole. As shown in Figure \ref{FE}, the $E_F^{\rm int}$ level is close to the CBM, and the formation energy at $E_F^{\rm int}$ for all electronic states is neutral or negative ionic ground state. Therefore, the population of Mu valence for the relaxed-excited states implies that the overall charge average is close to that of thermal equilibrium.  In other words, the reason why Mu (and H) simultaneously assumes donor and acceptor states in the relaxed-excited state is presumably to maintain the global charge neutrality of the system consisting of the host and H.

In conclusion, assuming that the electronic state of Mu implanted into MgH$_2$ takes the two pairs of defect centers (relaxed-excited states) associated with the donor/acceptor levels for interstitial Mu and HMu molecule states predicted from the DFT calculations, a scenario that coherently describe the present experimental results can be constructed.

\subsection{Implications to the kinetics and diffusion of interstitial H}
Now, let us discuss what we can learn for H kinetics from these results. It has been reported that the addition of a few \% of transition metals such as Ti significantly improves the H kinetics \cite{Liang:99,Varin:07,Amama:11}. DFT calculations suggest that, when a transition metal exhibits negative $U$ character to accompany a double charge conversion level $E^{+/-}_{\rm TM}$ in the host, the addition of the transition metal is expected to shift the Fermi level by $\Delta E=E^{+/-}_{\rm TM}-E_F^{\rm int}$ \cite{Roy:13}. As can be seen in Figure \ref{FE}(a), for $\Delta E>0$ (as predicted for Ti), the interstitial H is stable in the H$_{\rm i}^-$ state, and this is assumed to be the cause of the kinetic improvement by avoiding binding of H$_{\rm i}$ to the ligand H$^-$.
 
On the other hand, it is also reported that ball milling with a small amount of Nb$_2$O$_5$ improves the kinetics \cite{Hanada:07}. From $\mu$SR in the samples after milling with 5 wt\% Nb$_2$O$_5$, a decrease of the characteristic temperature $T_{\rm d}$ where diffusion motion of H and/or Mu sets in (as inferred from sharp decrease in $\Delta_{\rm KT}$) from 650 K to about 450 K was observed \cite{Umegaki:14}.  Assuming that Nb$_2$O$_5$ is an oxidation catalyst and that stabilization of the H$_{\rm i}^-$ state is the key to improving kinetics, it can be interpreted that Nb$_2$O$_5$ raises bulk $E_F$ by causing dehydrogenation; $${\rm MgH}_2 \rightarrow {\rm MgH}_{2-\delta}+\frac{\delta}{2}{\rm H_2}+\delta e^-.$$ The fact that a similar effect is obtained in the case of ball milling alone suggests that the increased surface area also promotes dehydrogenation reactions by atmospheric oxygen. It is paradoxical that dehydrogenation corresponds to {\sl reduction} for hydrides. 

Meanwhile, it has also been reported that the formation of the Mu$_{\rm 3S}$ state (corresponding to the H$^-$-H$^+$-H$^-$ complex state) is promoted in those samples \cite{Umegaki:14}. Since such hydrogen bond formation is expected to be rather unfavorable from a diffusion point of view, it is assumed that the reason why H kinetics improves with Nb$_2$O$_5$ addition is because the effect of stabilizing H$_{\rm i}^-$ outweighs that of hydrogen bond formation; the upward shift of $E_F$ will eventually lead to the dissociation reaction, $${\rm H}^-\mathchar`-{\rm H}_{\rm i}^+\mathchar`-{\rm H}_{\rm i}^- +2e^-\rightarrow {\rm H}^-+[{\rm H}_2]_{\rm i}^-+ e^-\rightarrow {\rm H}^-+2{\rm H}_{\rm i}^-.$$  A similar situation is suggested from an earlier $\mu$SR study for the effect of Ti-doping to NaAlH$_4$ \cite{Kadono:08,Hiraishi:arXiv} 

Regarding the Mu$^-_{\rm KT}$ state corresponding to H$_{\rm i}^-$, the magnitude of $\Delta_{\rm KT}$ is found to vary little over the entire temperature range, indicating that the Mu site remains unchanged over the time range of $\sim$10$^1$ $\mu$s. The fluctuation frequency $\nu_{\rm KT}$ shows nearly constant value of $\sim$0.1 MHz below 150 K, which is unlikely to be attributed to thermally activated diffusive motion of Mu$_{\rm KT}^-$. The non-monotonic temperature dependence of $\nu_{\rm KT}$ above 150 K also suggests that the cause is non-thermal.  In addition, the paramagnetic Mu components (Mu$_{\rm p}$/Mu$_{\rm d}$ and Mu$_{\rm p2}$) show fast spin and/or charge exchange reactions with excited carriers even below 150 K. 
Considering that the frequency of paramagnetic fluctuations ($>10^2$ MHz) are far from the sensitive range for the diamagnetic Mu, these observations suggest that the fluctuations of $\Delta_{\rm KT}$ is induced by the relaxation of proton spins of the nearby H('s) in the relaxed-excited states (denoted as H$^*$) generated upon muon implantation (see Figure \ref{model2}).

Note that H$^*$ can be any of the observed four Mu states with H substituting for Mu, which may be appropriately referred to H$_{\rm p}$/H$_{\rm d}$, H$_{\rm KT}$,  H$_{\rm p2}$, and H$_{\rm 3S}$. The proton spin relaxation rates $\lambda^*$ in these states can be evaluated by reducing those for the diamagnetic Mu (inferred from Figure \ref{params}) by a factor $R^2\equiv[\gamma_{\rm p}/\gamma_{\rm \mu}]^2$ ($=0.0986$) for the same fluctuation frequency as observed in Mu, e.g., $\lambda^*=R^2\lambda_{\rm 3S}\simeq0.2$--0.5 MHz for the H$_{\rm 3S}$ state.  This implies that further experimental verification will be necessary to determine whether the fluctuations of nuclear dipolar fields exerted by H to the Mg atom binding a negative muon (pseudo-Na atom) observed by negative muon $\mu$SR \cite{Sugiyama:18} can be solely attributed to H diffusion, because the pseudo-Na atoms can accompany nearby H$^*$'s due to local electronic excitations which are much stronger than that in positive muon implantation \cite{Okumura:21}.

\section*{Acknowledgement}
We acknowledge helpful discussions with J. Sugiyama, I. Umegaki, and H. Kageyama. The $\mu$SR experiments were conducted
under the Inter-University Research Program (Proposal No.~2019MS02) supported by the Institute of Materials Structure Science, High Energy Accelerator Research Organization (KEK), Japan. This work was supported by the MEXT Elements Strategy Initiative to Form Core Research Center for Electron Materials (Grant No. JPMXP0112101001), JSPS KAKENHI (Grant Nos.~19K15033, 17H06153, 20K12484,  21H05102), Core-to-Core Program (JPJSCCA20180006), and JST MIRAI Program (JPMJMI21E9). 
\vspace{5pt}

\noindent
Data Availability: The data that supports the findings of this study are available within the article.

\vspace{0.5cm}

\setcounter{figure}{0}
\setcounter{table}{0}
\setcounter{equation}{0}
\renewcommand{\thefigure}{A\arabic{figure}}
\renewcommand{\thetable}{A\arabic{table}}
\renewcommand{\theequation}{A\arabic{equation}}

\section*{APPENDIX A: Details of \msr\ data analysis}
The spin relaxation of the diamagnetic Mu is governed by a random local fields at the Mu site ${\bm H}_{\rm n}$ from the nuclear magnetic moments, and the magnitude of the relaxation rate is determined by the second moment of ${\bm H}_{\rm n}$,
\begin{equation}
\Delta^2=\gamma_\mu^2\langle|{\bm H}_{\rm n}|^2\rangle=\gamma_\mu^2\sum_n f_n\sum_i\sum_{\alpha,\beta}[\gamma_n(\hat{A}_{\rm d}^{ni})^{\alpha\beta}\overline{{\bm I}}_{ni}]^2, \label{delta_n}
\end{equation}
where $\gamma_\mu=2\pi \times 135.53$ MHz/T is the gyromagnetic ratio of Mu spin, ${\bm I}_{ni}$ is the $n$th kind of nuclear spin at distance $r_{ni}$ ($n=1,2$ for $^{1}$H, $^{25}$Mg) on the $i$th lattice point (with $\gamma_n\overline{{\bm I}}_{ni}$ being the effective magnetic moment considering the electric quadrupole interaction for ${\bm I}_{ni}\ge1$), $\gamma_n$ and $f_n$ are the gyromagnetic ratio and natural abundance (or occupation) of the $n$th nuclear spin, $\hat{A}_{\rm d}^{ni}$ is the magnetic dipole tensor
\begin{equation} 
(\hat{A}_{\rm d}^{ni})^{\alpha\beta}
=\frac{1}{r^3_{ni}}(\frac{3\alpha_{ni} \beta_{ni}}{r^2_{ni}}-\delta_{\alpha\beta}) \:\:(\alpha,\beta=x,y,z),\label{diptensor}
\end{equation} 
representing the hyperfine interaction between Mu and nuclear magnetic moments at  ${\bm r}_{ni} = (x_{ni},y_{ni},z_{ni})$ being the position vector of the nucleus seen from Mu.  The sum in equation (\ref{delta_n}) takes all $x,y,z$ for $\beta$, and $x,y$ for the $\alpha$ components that are effective for longitudinal relaxation when $\hat{z}$ is the LF direction; the $z$ component does not contribute to the relaxation because it gives a magnetic field parallel to the Mu spin. 

When the coordination of the nuclear magnetic moment viewed from Mu is isotropic and the number of coordination $N$ is sufficiently large ($N\ge4$), the time-dependent spin relaxation under zero external field is described by the Gaussian Kubo-Toyabe (GKT) function
\begin{equation}
G_z^{\rm KT}(t;\Delta) = \frac{1}{3}+\frac{2}{3}(1-\Delta^{2}t^{2})e^{-\frac{1}{2}\Delta^{2}t^{2}}. \label{gkt}
\end{equation}
In the case of fluctuating $\Delta$ over time (e.g., due to the self-diffusion of Mu), $G_z^{\rm KT}(t; \Delta)$ is subject to the modulation and adiabatically approaches exponential decay with increasing fluctuation rate $\nu$, where the detailed lineshape of $G_z^{\rm KT}(t;B_0,\Delta,\nu)$ as a function of $\nu$ and LF ($=B_0$) are found elsewhere \cite{Hayano:79}.  For a small number of nuclei, the time evolution of the Mu spin polarization exhibits considerable deviation form the GKT function. The exact time dependence can be derived from the density matrix of the Mu-nucleus spin system 
for the $N=1$ and 2 cases \cite{Brewer:86,Nishiyama:03,Wilkinson:20} and the latter is introduced to
describe H-Mu-H hydrogen bonding state \cite{Kadono:08,Umegaki:14}.
Assuming a static collinear geometry with $\mu^+$ at the center of the line joining two other nuclear spins ($I=1/2$), the time evolution of 
muon polarization as a cubic average is calculated by solving a simple 
three-spin model to yield \cite{Brewer:86}
\begin{eqnarray}
G_z^{\rm 3S}(t)&=&\frac{1}{6}\{3+\cos(\sqrt{3}\omega_{\rm 3S}t)+\alpha_{+}
\cos(\beta_{+}\omega_{\rm 3S}t) \nonumber\\
& &\:\:+  \alpha_{-}\cos(\beta_{-}\omega_{\rm 3S}t)\}e^{-\lambda_{\rm 3S}t},\label{G3S}
\end{eqnarray}
where $\alpha_\pm=1\pm1/\sqrt{3}$, $\beta_\pm=(3\pm\sqrt{3})/2$, $\omega_{\rm 3S}$ is the dipolar interaction frequency
\begin{equation}
\omega_{\rm 3S}=\gamma_\mu\gamma_1/r_{\rm t}^3, \label{omgd}
\end{equation}
with $r_{\rm t}$ being the distance between $\mu^+$ and the proton nucleus, $\gamma_1/2\pi=42.58$ MHz/T for $^1$H,  and $\lambda_{\rm 3S}$ is the relaxation rate describing the influence of local fields from outer-shell nuclear magnetic moments \cite{Wilkinson:20} and spin/charge exchange interaction.  

In the case of fast fluctuation ($\nu>\Delta$), we have an approximated form of the GKT function
\begin{equation}
G_z^{\rm KT}(t; B_0, \Delta, \nu)\simeq\exp(-\lambda t),\label{Gexp}
\end{equation}
where
\begin{equation}
\lambda\simeq\frac{2\Delta^2\nu}{\omega^2+\nu^2},\label{BPP}
\end{equation}
and $\omega =\gamma_\mu B_0$. Note that when the fluctuation rate satisfies the condition $\nu\gg\omega$, equation (\ref{BPP}) is least dependent on $B_0$. 

The magnitude of $\Delta$ is sensitive to the size of the nearest-neighbor nuclear magnetic moment $\gamma_n\overline{{\bm I}}_{ni}$ and the distance $|{\bm r}_{ni}|$ from Mu, and the position occupied by Mu as pseudo-H can be estimated by comparing the experimentally obtained $\Delta$ with the calculated value at the candidate sites. We used the DipElec code \cite{Kojima:04} to calculate $\Delta$ for MgH$_2$.

The $\mu$SR time spectrum shown by the paramagnetic Mu is complicated by the hyperfine interaction between $\mu^+$ and the bound electron. In addition to the limited time resolution of pulsed beam experiments, it is difficult to directly observe the Mu spin precession motion associated with transitions between hyperfine levels due to the dominance of fast spin/charge exchange reactions in MgH$_2$. Even in such a case, important information can be extracted by examining the response of the time spectrum $G_z(t)$ to the LF in detail.

First, consider Mu$^0$ under a magnetic field ($B_0$) applied parallel the initial Mu polarization ($\parallel z$) without spin fluctuations (static).  In this case, we have
\begin{equation}
G_z(t)=\frac{1}{2(1+x^2)}\left[(1+2x^2)+\cos(\omega_c\sqrt{1+x^2})t\right],\label{MuPz}
\end{equation}
where $x=B_0/B_c$ with $B_c=\hbar\omega_c/(\gamma_\mu-\gamma_e)$ being the effective local field exerted on $\mu^+$, $\omega_c$ is the hyperfine parameter (angular frequency for the isotropic Fermi contact term), and $\gamma_e$ is the electron gyromagnetic ratio ($=2\pi\times28024.2$ MHz/T) \cite{Patterson:88}.
For the zero external field ($x=0$), 
\begin{equation}
G_z(t)=\frac{1}{2}(1+\cos\omega_c t).\label{MuPz2}
\end{equation}
The implication is that a half of the ground state is not an eigenstate of the hyperfine interaction, so the corresponding component exhibits oscillation between the spin-singlet and triplet states with $\omega_c$.  In a practical material, $\omega_c$ is often smaller than that in a vacuum ($\omega_{\rm vac}=2\pi\times4463.3$ MHz) due to the effect of dielectric constant, etc., but it is often not visible in ordinary time-resolution measurements (i.e., the second term is zero when time-averaged). Thus, half of the implanted muons effectively lose their polarization upon Mu$^0$ formation (also called the missing fraction), and only the triplet state [the first term of equation (\ref{MuPz})] becomes observable. The situation is similar for $B_0>0$, where the time-averaged polarization at $t=0$ is
\begin{equation}
G_z(t)\simeq\overline{G}_z(0)=\frac{1+2x^2}{2(1+x^2)}\equiv g_z(x).\label{MuPzr}
\end{equation}
As $x$ is normalized by $\omega_c$, the field dependence of $\overline{G}_z(0)$ varies with $\omega_c$. Moreover, Mu$^0$ in MgH$_2$ is subjected to the NHF interaction between the bound electron and surrounding H atoms which have proton nuclear spins \cite{Patterson:88}. This further reduces $\overline{G}_z(0)$ to $1/2\times1/3=1/6$, reflecting the random orientation of the nuclear spins, and leads to a two-step recovery of $\overline{G}_z(0)$ with respect to $x$ \cite{Beck:75}.  In any case, the magnitude of $\omega_c$ can be determined by comparing the magnetic field dependence of experimentally obtained $\overline{G}_z(0)$ with equation (\ref{MuPzr}). 

The effect of spin fluctuations on Mu$^0$ under LF can be derived by approximating the spin/charge exchange reaction with excited carriers by a strong collision model. In particular, when the reaction rate $\nu$ is lower than $\omega_c$, the time-dependent polarization is given by the following equations \cite{Kadono:03}.
\begin{equation}
G_z(t)\simeq g_z(x)\exp(-\lambda t),\label{omup}
\end{equation}
\begin{equation}
\lambda\simeq \frac{\nu}{2(1+x^2)}.\label{rlx}
\end{equation}

\begin{figure}[t]
\begin{center}
\includegraphics[width=0.7\linewidth]{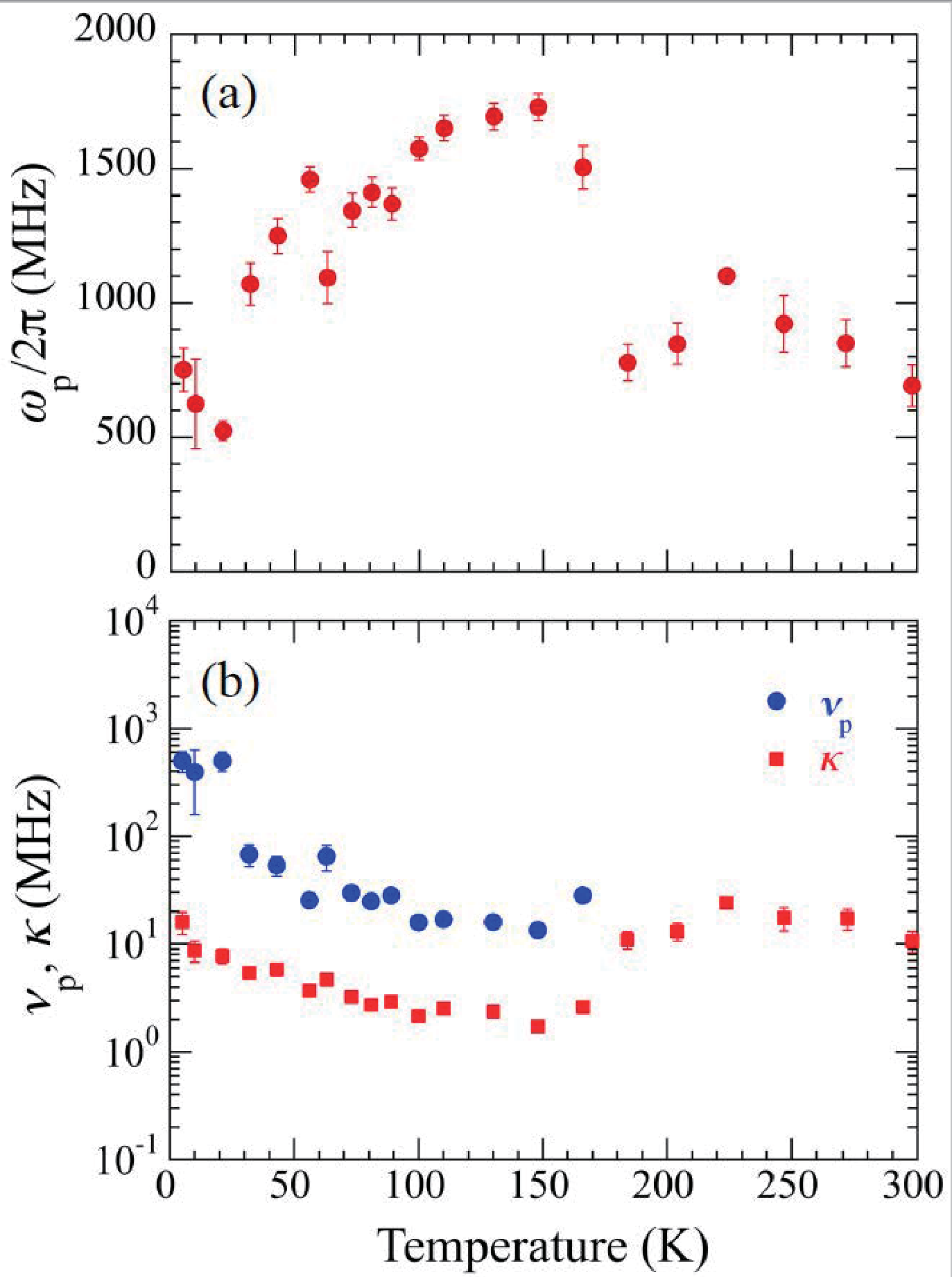}
\caption{Temperature dependence of (a) the hyperfine parameter ($\omega_{\rm p}$), (b) spin/charge exchange rate ($\nu_{\rm p}$) and conversion rate ($\kappa$) for the Mu$_{\rm p}$ state obtained in the first-round analysis in which $\omega_{\rm p}$ was allowed to vary as a free parameter, where $\nu_{\rm p}$ was hardly determined above $\sim$150 K.}
\label{omgp}
\end{center}
\end{figure}

As mentioned in the main text, we attempted to derive the hyperfine parameter ($\omega_{\rm p}$) of a paramagnetic state (Mu$_{\rm p}$) and other kinetics parameters describing its dynamical properties by using equation (\ref{gp}) [whose magnetic field dependence originates from equations (\ref{MuPzr})--(\ref{rlx})] by a global curve fit that simultaneously analyzes the spectra of different magnetic fields. As a result, $\omega_{\rm p}$ increases significantly from around 20 K to 150 K, and shows a sharp decrease at the high temperature side above 150 K [Figure \ref{omgp}(a)]. Moreover, the latter is strongly correlated with the increase of the spin/charge fluctuation rate $\nu_{\rm p}$, leaving it hardly determined above $\sim$150 K [Figure \ref{omgp}(b)]. Therefore, we adopted the hypothesis that the original $\omega_{\rm p}$ should be approximately constant, and derived the temperature dependence of the other parameters with $\omega_{\rm p}$ fixed to some appropriate value. The temperature dependence of $\omega_{\rm p}$ ($\omega_{\rm p2}$) and its relationship to $\nu_{\rm p}$ ($\nu_{\rm p2}$) are discussed in the main text.

\setcounter{figure}{0}
\setcounter{table}{0}
\setcounter{equation}{0}
\renewcommand{\thefigure}{B\arabic{figure}}
\renewcommand{\thetable}{B\arabic{table}}
\renewcommand{\theequation}{B\arabic{equation}}

\section*{APPENDIX B: Muon irradiation effect}

In general, the energy imparted in a solid by the irradiation of charged particles consists of two types of energy: the kinetic energy of the atoms that make up the solid and the excitation energy of the electron system. The former is due to elastic collisions between incident particles and atoms, and the latter is due to inelastic collisions. Irradiation damage (atomic displacement), usually seen in metals and semiconductors, occurs when atoms gain energy above the knock-on threshold ($E_{\rm d}\simeq15$--30 eV) and are ejected from a lattice point \cite{Thompson:74}. Provided that the mass of the incident particle is $m$, the energy is $E$, and the mass of the target atom is $M$, ejection occurs when the energy $\Delta E$ imparted to the atom by the collision exceeds $E_{\rm d}$, i.e.,
\begin{equation}
\Delta E=\frac{4mME}{(m+M)^2}>E_{\rm d}.
\end{equation}
When the incident particle is a muon ($m= m_\mu\ll M$), the minimum kinetic energy $E_0$ required to eject the target atom is given by
\begin{equation}
E_0=\frac{(m+M)^2}{4mM}E_{\rm d}\simeq\frac{M}{4m_\mu}E_{\rm d}.
\end{equation}
Therefore, it is inferred that in the region $E>E_0$, muon kinetic energy is consumed mainly by the atomic displacement process.

On the other hand, in the region of $E\le E_0$ (i.e., $\Delta E\le E_{\rm d}$), the energy loss due to excitation of the electronic system becomes dominant. Since the energy required for the formation of electron-hole pairs is known to be approximately $\varepsilon\simeq2.8E_{\rm g}$ with the band gap as $E_{\rm g}$ \cite{Alig:75}, the number of electron-hole pairs produced is roughly estimated as 
\begin{equation}
n_{eh}\simeq \frac{E_0}{\varepsilon}=\frac{E_0}{2.8E_{\rm g}}.
\end{equation}
Assuming $E_{\rm d}=30$ eV and $E_{\rm g}=5$ eV, the values of $E_0$ and $n_{eh}$ estimated for the constituent atoms of MgH$_2$ are summarized in Table \ref{Exciton}. From the weighted average of these values, it is estimated that about 10$^2$ electron-hole pairs per muon are formed in MgH$_2$. Since such electronic excitations are dominant in the final stage of energy loss of the incident particles, electron-hole pairs are also expected to be distributed near the end of the radiation track.

From the above discussion, H is special in that muons with relatively low kinetic energy are inferred to displace H atoms in solids. Considering that the formation energy of H vacancies in MgH$_2$ is predicted to be comparable with that for the interstitial H (H$_{\rm i}$) \cite{Park:09}, it is possible that the ejected H exists as H$_{\rm i}$ in the vicinity of the muon stopping position.

\begin{table}[h]
\caption{The minimum kinetic energy $E_0$ of the incident muon required for each target atom to undergo displacement in MgH$_2$, and the number of electron-hole pairs $n_{eh}$ produced by electronic excitation below $E_0$, where $E_{\rm d}= 30$ eV and $E_{\rm g}= 5$ eV.}\label{Exciton}
\begin{center}
  \renewcommand\arraystretch{1.2}
\footnotesize
\begin{tabular}{ccc}
\br
\hspace{2mm}Atom\hspace{2mm} & \hspace{2mm}$E_0$\hspace{2mm} (eV) & \hspace{2mm}$n_{eh}$\hspace{2mm}\\
\mr
H & 67 & 4.8 \\
Mg & 1619 & 116\\
\br
\end{tabular}
\end{center}
\end{table}
\def\urlprefix{}
\def\issnprefix{}
\section*{References}
\input{MgH2.bbl}

\end{document}

%% file: MgH2.bbl
\providecommand{\newblock}{}